\documentclass{ws-ijmpa}

\begin{document}

\markboth{R.Delbourgo and P.D.Stack}
{GR of space-time-property}
\catchline{}{}{}{}{}

\title{Where-When-What: the General Relativity of Space-Time-Property}

\author{Robert Delbourgo and Paul D Stack}

\address{School of Mathematics and Physics, University of Tasmania, Locked Bag 37 GPO\\ Hobart, Tasmania 7001,AUSTRALIA 7001\\
bob.delbourgo@utas.edu.au, pdstack@utas.edu.au}

\maketitle

\begin{abstract}
We develop the general relativity of extended spacetime-property for
describing events including their properties. The anticommuting nature of 
property coordinates, augmenting space-time $({\bf x},t)$,
allows for the natural emergence of generations and for the simple incorporation 
of gauge fields in the spacetime-property sector.
With one electric property this results in a geometrical unification of gravity
and electromagnetism, leading to a Maxwell-Einstein Lagrangian plus a
cosmological term. Addition of one neutrinic and three chromic properties should 
lead to unification of gravity with electroweak and strong interactions.
\end{abstract}

\maketitle

\section{\label{sec1}A full description of events}

In a static universe nothing happens. All systems continue inertially, never change
and never interact; that sort of universe is the ultimate non-event. Even speaking 
of an `observer' is a contradiction in terms, because every observer would be
incommunicado and unaware of anything and everything. So, more to the point,
a static universe is purely hypothetical and logically inconceivable. On the  contrary
the real universe is in a state of flux. It is punctuated by a succession 
of events from which we gain the notion of space and time, as the scenario 
unfolds. Historically the spacetime arena has held centre stage ever 
since the ideas of relativity took hold about a hundred years ago and we are
accustomed to characterising an event by its spacetime location, even to
the extent of describing events geometrically. Indeed the geometry of curved
spacetime has revolutionised our ideas about gravity ever since Einstein's
development of general relativity.

When an event occurs, it amounts to a change in circumstances, whereby
an object alters its motion and possibly character as it engages with others.
(It is the impact of these changes in spacetime which underpins the process
of observation and the Heisenberg uncertainty principle.)
Thus when a photon is emitted and reabsorbed by two charged objects we
interpret this as a succession of two events resulting in an electric interaction 
between the charged objects. And since quantum field theory came into being,
we recognise this through a trilinear interaction between the charged object and a 
photon with the conservation of total energy-momentum at the event vertex, which
is ensured by taking an integral over all spacetime of the interacting fields. So 
much is well understood. However saying that an event has happened {\em there} 
and {\em then} does not fully specify the event; we must in addition specify 
{\em what} exactly has happened: what sort of transaction has taken place and
what properties may have been exchanged. For instance when a proton emits a 
neutron and charged pion virtually we have to add that information
and then the event becomes fully explicated. This is normally done by
introducing quantum fields with particular labels and interacting in a manner
that usually conserves some quantum number, such as isospin for the nuclear
example. 

As physicists, we are well aware that particle properties do proliferate 
but they can be systematised using group theory of some ``internal group'', 
resulting in certain types of representations of various Lie algebras. These
can be constructed from some fundamental representation as a result of
some basic dynamics connected with primitive constituents like quarks and
leptons. As new particles are discovered it occasionally becomes necessary to
enlarge the group to accommodate features or properties that do not
fall comfortably within the conventional picture; this is how progress in quantum 
field theory has been achieved and it has led to its ultimate incarnation as the
standard model, wherein U(1)$\times$SU(2)$\times$SU(3) reigns supreme. 
Nevertheless some issues remain unresolved, such as the generation problem, 
where repetitions of particle families lead to further concepts about ``horizontal 
groups'' for systematising them -- although what the correct group and 
representations are is still unsettled. Most worrying of all is the number of 
parameters needed to characterise the interactions and masses of the
(light) three generations in the standard model, though it must be admitted that 
those parameters are sufficient to describe a {\em vast} number of experimental 
facts. This has stimulated research into looking for some kind of grand unification 
for reducing the number of parameters and it has spawned a large number 
of interesting new concepts. Supersymmetry and superstrings, based on
enlarged dimensions, are the foremost amongst these as they automatically 
solve the fine-tuning problem and naturally incorporate gravity, but we should 
not discount other ideas such as technicolour, preons, non-commuting 
spacetime variables, deformation groups, and so on.

Underlying all these extensions, the question arises: how to enumerate properties 
and characteristics of events at a fundamental particulate level. Traditionally one 
specifies the associated quantum fields by attaching as many labels to them as 
needed to describe the properties and by ensuring that the events due to 
their mutual interactions conserve whatever properties remain intact overall; 
in other words adapting to what the experiments dictate. This approach is 
predictive to the extent that if there are any missing components of the group 
representations, then they must eventually show up experimentally and
conform to the overarching symmetry. And if the interactions do not preserve 
the expected symmetries, they are nowadays thought to be due to symmetry 
breaking effects coming from a yet unknown cause, possibly spontaneously 
generated.

In an effort to characterise the nature of an event, we have, over a
number of years, suggested that it may be possible to specify attributes or
`properties' of particles by connecting them with a few anticommuting Lorentz
{\em scalar} coordinates (and their conjugates or opposites). An event can
thereby be fully described by a conglomerate there-then-what label; by analogy
to momentum conservation, the conservation of overall property, such as 
charge, is guaranteed by integrating appropriately over all property space. 
To that end we have suggested that the spacetime manifold with coordinates 
$x$, should be enlarged to a supermanifold \cite{BD,BDW,FB,AR,EW} by attaching
a few complex anticommuting $\zeta$ coordinates, with fields being regarded 
as functions of $X=(x,\zeta,\bar{\zeta})$. The anticommuting character
of fermionic quantum fields is nothing new and the use of Grassmann
variables has featured in the BRST quantisation of gauge theories, and
more especially in the invention of supersymmetry -- although the latter
makes them Lorentz spinor rather than scalar. Once a property is ascribed
it cannot be doubly ascribed, corresponding to the fact the square of
a given Grassmann variable vanishes. But because the conjugate property is
available, one can built up property scalars that may be multiplied with
other properties, and so on. This is how one might conceive of families of 
systems with similar net attributes \cite{RDPJRW,RDRZ}. Furthermore the 
anticommuting character of a fixed number of properties means that we can never
encounter an infinite number of states, unlike bosonic extra dimensional
spaces modelled along Kaluza-Klein lines. The mathematics of graded spaces 
is quite well understood now and we will take full advantage of this in what 
follows.

Previous investigations have shown that with a minimum of five independent
$\zeta$ one can readily accommodate all the known particles and their
{\em generations}. We might tentatively call these properties: charge or 
electricity, neutrinicity and chromicity. The use of five is no accident, being
based solidly on economical grand unified groups such as SU(5) and SO(10); 
what is more interesting is that the correct representations of those algebras 
come out automatically. [In our case Sp(10) is the overarching internal group.]
Since addition of Grassmann variables has the mathematical effect of reducing 
the {\em net} `dimension' of the space, there is the tantalising 
prospect of a universe with zero total dimension. We have elaborated on 
some consequences of such a scheme \cite{RD2,RD3}, including the appearance 
of a few new particles and the possibility of reducing the number of parameters 
appearing in the standard model, because we have just nine Higgs fields but 
only one coupling constant attached to all known fermions \cite{RD4}.

The purpose of the present paper is to describe the general relativity of 
spacetime-property along the lines originally devised by Einstein.
Our aim is quite modest: we wish to see if one can unite electromagnetism
with general relativity geometrically in a way which differs from
Klein-Kaluza and (spinorial) supergravity in as much as the extended
coordinates are {\em scalar} but anticommuting. Having established that, it
means one may contemplate generalizations which include other forces, 
without producing  infinite towers of excited states. This paper should
be seen as a first step in that direction; further avenues for research are
mentioned in the conclusions and include Higgs fields and possibly ghost
fields associated with gauge-fixing at the quantized level.
The extended metric will consist a $x-x$ sector, a $\zeta-x$ sector and
a $\zeta-\bar{\zeta}$ sector; the space-space sector involves the gravitational
field as well as some extra $\bar{\zeta}\zeta$ terms, the space-property 
sector brings in the gauge fields connected with property propagation,
and the curved property-property sector can be the source of the cosmological 
term, as it happens; it is no surprise that the communicators of property or 
gauge fields should reside where they are, but the association of property
curvature with a cosmological term is perhaps more intriguing. It is important
to state that fermion fields, such as the gravitino, have no place in the extended
metric as they would cause a conflict with Lorentz invariance. Because the 
fermionic superfield $\Psi_\alpha$ carries spinor indices, it must be studied
 separately and in that regard our approach differs radically from conventional
 spinorial  supersymmetry; we give a preliminary treatment of fermions
 towards the end of the article.

As we are dealing with a graded manifold or supermanifold, it is {\em essential} to 
make sure that the super-coordinate transformations carry the correct commutation 
factors. This is carefully explained in Sections 2 to 5, where our conventions are 
established and summarized to make the paper self-contained. A good notation is of 
course vital {\cite{RDPDJGT,RDPDJGT2} and the ordinary general relativity convention with its 
placement of indices conflicts with conventional differentiation as a left 
operation (as a rule for the average reader); we have had to compromise on that 
as some traditional ideas are not easily overthrown. Asorey and Lavrov \cite{MAPL} 
have written a nice exposition of these ideas but they have instead chosen to take 
right derivatives, which makes for marginally simpler formulae but does not conform 
with traditional ideas about left differentiation, which we have religiously adhered to. 
In Section 3 we delineate transformation properties of supertensors and the 
supermetric. Then we
pay attention to the definition of covariant derivatives  with particular application
to the super-Riemann tensor ${\cal R}$; its supersymmetry properties are obtained
and the super-Ricci and superscalar curvature are derived. The Bianchi
identities follow in Section 5. For the remainder of the paper we consider the case 
of just one property, such as charge, and in Section 6 we write down the most 
general metric, including the electromagnetic field. Unsurprisingly we
show that gauge transformations can be construed as supercoordinate
transformations associated with phase changes on $\zeta$. We then find the 
superdeterminant of this metric in Section 7, as well as the Palatini form of the 
superscalar curvature.  
In Section 8 we go on to evaluate the super-Riemann tensor components and determine
the super-Ricci tensor and full supercalar curvature. This leads to the field equations 
and it is rather pleasing that the Einstein-Maxwell Lagrangian emerges very 
naturally, together with a cosmological contribution. Further, the electromagnetic
stress tensor presents itself as a purely geometrical addition to the extended
Einstein tensor. All such calculations are greatly assisted by an algebraic 
computer program for handling anticommuting variables as well as ordinary 
ungraded ones, which has been developed by one of us (PDS) using
{\em Mathematica}. A penultimate Section 9 details the inclusion of matter fields, 
but our treatment of fermions there is to be regarded as preliminary at this stage. 
The conclusions close the paper and an Appendix collates a list of generalised 
Christoffel symbols, curvature components and super-vielbeins that are needed 
at intermediate steps in the main text.

\section{\label{sec2}Extended transformations and notation}
The addition of extra anticommutative coordinates to space time results 
in a graded manifold, where the standard spacetime is even and the 
property sector is odd. The notation used in this work will be to define 
uppercase Roman indices ($M$, $N$, $L$, etc) to run over all the dimensions 
of spacetime-property and hence have mixed grading. Lower case roman 
indices ($m$, $n$, $l$, etc) will correspond to even graded spacetime, and 
Greek characters ($\mu$, $\nu$, $\lambda$, etc) will correspond to the odd 
graded property sector. The grading of an index given by [$M$] \footnote{Asorey 
and Lavrov use the notation $\epsilon_M$ instead of $[M]$.}, 
is $[m] = 0$ and $[\mu] = 1$.  Later on we reserve early letters of the 
alphabet (a,$\alpha$, etc.) to signify flat or tangent space.

Our starting point is the transformation properties of contravariant and 
covariant vectors; from these we can build up how a general tensor should 
transform. We will make use of Einstein summation convention but it has to 
be done carefully. We pick a convention of always summing a contravariant 
index followed immediately by a covariant index (up then down). This results 
in contravariant and covariant vectors transforming as follows:
\begin{equation} \label{vectortrans}
V^{\prime M}=V^N\frac{\partial X^{\prime  M}}{\partial X^N}, \;\;\;  
V^{\prime}_M=\frac{\partial X^N}{\partial X^{\prime  M}}V_N.
\end{equation}

The scalar $V^M V_M$ then correctly transforms into itself: 
\begin{equation} \label{scalartrans}
V^{\prime M} V^{\prime}_M=V^N\frac{\partial X^{\prime  M}}{\partial X^N}
\frac{\partial X^L}{\partial X^{\prime  M}}V_L=V^N \delta_N{}^L V_L =  V^N V_N,
\end{equation}
since the (left) chain rule given by
\begin{equation}\label{chainrule}
\frac{\partial X^{\prime  M}}{\partial X^N}\frac{\partial X^L}
{\partial X^{\prime  M}} = \delta_N{}^L.
\end{equation}
From (\ref{vectortrans}) one can build up the transformation properties of any tensor, 
by taking it to behave like a corresponding product of vectors. For example 
a rank two covariant tensor $T_{MN}$ has to transform like $V_M V_N$.
\begin{eqnarray}
 V^{\prime}_M V^{\prime}_N &=&\frac{\partial X^R}{\partial X^{\prime  M}}
 V_R \frac{\partial X^S}{\partial X^{\prime  N}}V_S  \nonumber \\
&=& (-1)^{[R]([S]+[N])} \frac{\partial X^R}{\partial X^{\prime  M}} 
\frac{\partial X^S}{\partial X^{\prime  N}}V_R V_S. 
\label{2covtrans} \end{eqnarray}
In these manipulations we have adhered to the traditional convention
of writing derivatives on the left, so the the sign factor arising in (\ref{2covtrans}) is due to 
permuting $V_R$ through the partial derivative. In this way we find how
$T_{MN}$ transforms:
\begin{equation}\label{tensortrans}
T^{\prime}_{MN} = (-1)^{[R]([S]+[N])} \frac{\partial X^R}{\partial X^{\prime  M}} 
\frac{\partial X^S}{\partial X^{\prime  N}}T_{RS}.
\end{equation}
Thus in (\ref{tensortrans}) we do not have an immediate, direct up-down summation; 
the sign factor is introduced to compensate for this. In this manner it is not
hard to derive sign factors for any sort of tensor.

\section{\label{sec3}On metrics and supertensors}
The metric supertensor $G_{MN}$ is chosen to be graded symmetric, 
$G_{MN} = (-1)^{[M][N]} G_{NM}$ because it is associated with the
generalised spacetime-property separation $ds^2= dX^N dX^M G_{MN}$,
which is overall  bosonic.
As with standard general relativity the metric can be used
to raise and lower indices; however the direct up-down summation 
rule must be strictly obeyed. This means that for vectors \footnote{Note that 
in ref.7 the $G$ with raised indices differs from the present $G$
by a factor $(-1)^{[N]}$.}:
\begin{equation} \label{raiselowervectormetric}
V^M G_{MN} = V_N, \;\;\; G^{MN} V_N = V^M
\end{equation} 
If this order is not followed then the resulting vector (or tensor) will not 
transform correctly according to the rules in Section \ref{sec2}. When raising 
or lowering indices of a supertensor an adjoining up-down summation is 
sometimes impossible; in that event a sign factor like in (\ref{2covtrans}) must again be 
included to compensate for this, using the same argument. 

For illustration, consider a tensor $T_{MN}$ whose second index $N$ 
we wish to raise to get $T_{M}{}^{N}$. To work out the sign factor required,
look at a product of vectors instead, say $T_{MN} = U_M V_N$; then
\begin{equation} \label{raisesecondindex}
T_{M}{}^{N} = U_M V^N = U_M G^{NL} V_L
= (-1)^{[M]([N]+[L])} G^{NL} T_{ML}
\end{equation}
The sign factor ensures that $(-1)^{[M]([N]+[L])} G^{NL} T_{ML}$ 
behaves like $T_M{}^N$. This procedure extends to any tensor.

The inverse metric multiplies the covariant metric as follows:
\begin{eqnarray}
G^{MN} G_{NL} = {\delta^M}_L =(-1)^{[M]}{\delta_L}^M \label{metriccon1}\\ 
(-1)^{[N]} G_{MN} G^{NL} = \delta_M{}^L \label{metriccon2}
\end{eqnarray}
These equations are consistent with each other, (the metric and its inverse 
being graded symmetric) and with the transformation properties of the singlet 
$\delta_M{}^L$.They are also consistent with Asorey and Lavrov \cite{MAPL}.
In particular notice from (\ref{metriccon1}) that the trace operation introduces a negative 
sign where the fermionic sector is concerned, as is well-known.

\section{\label{sec4}Covariant derivatives and the Riemann supertensor}
The connection coefficients of standard general relativity in the case of zero 
torsion are defined to be :
\begin{equation} \label{standardgamma}
\Gamma{}_{mn}{}^k = \left( g_{lm,n}+g_{ln,m}-g_{mn,l} \right) g^{lk}/2
\end{equation}
We take this as the starting point for our covariant derivative, extending it to 
a graded manifold by allowing for sign factors.
\begin{equation} \label{testgamma}
\Gamma_{MN}{}^K= \big((-1)^{X_{LMN}} G_{LM,N} +
 (-1)^{Y_{LNM}} G_{LN,M}  - (-1)^{Z_{MNL}} G_{MN,L} \big)G^{LK}/2,
\end{equation}
where $X_{LMN}, Y_{LNM}$ and $Z_{MNL}$ need to be determined 
so as to guarantee that the covariant derivative of a covariant vector
transforms correctly as a rank 2 covariant tensor. We write the covariant 
derivative in semicolon notation as
\begin{equation}  \label{testcovderiv}
A_{M;N} = (-1)^{W_{MN}} A_{M,N} - \Gamma_{MN}{}^K A_K,
\end{equation}
where again $W_{MN}$ is a sign factor to be found. Expanding this out and 
finding the conditions on the signs  so that all second derivatives cancel 
and the remaining terms transform as a rank 2 covariant tensor we arrive at
\begin{equation}\label{covderivcovvector}
A_{M;N} = (-1)^{[M][N]}A_{M,N} - \Gamma_{MN}{}^K A_K,
\end{equation}
with 
\begin{equation}\label{gammaformula}
 \Gamma_{M\!N}{}^K \!\!=\!\big[ (-1)^{[M][N]\!+\![L]} G_{ML,N}
  + (-1)^{[L]} G_{NL,M}- (\!-1)^{[L][M]\!+\![L][N]+[L]} G_{MN,L} \big] G^{LK}/2.
\end{equation}
In a similar manner one may establish that
\begin{equation} \label{covderivconvector}
{A^M}_{;N} =(-1)^{[M][N]}({A^M}_{,N} + A^L{\Gamma_{LN}}^M)
 \qquad\qquad{\rm and}
\end{equation}
\begin{equation}\label{covderivcovtensor}
{T_{LM}}_{;N}=(-1)^{[N]([L]+[M])}[T_{LM,N}-{\Gamma_{NL}}^K T_{KM} 
 -(-1)^{[L]([M]+[K])}{\Gamma_{NM}}^KT_{LK}].
\end{equation}
The curious factors of $(-1)^{[M][N]}$, etc. arise from the mismatch between
left derivatives clashing with the convention of placing subscript such as
$_{,M}$ on the right and we are stuck with this inappropriateness. Anyhow,
with these constructions of the covariant derivative, it is pleasing to
check that $G_{MN;L}$ vanishes, as expected. And from equations such
as (\ref{covderivcovvector}),(\ref{covderivconvector}),(\ref{covderivcovtensor}), one may deduce the rules for covariant derivatives
of any supertensor \footnote{In this context, the rule for covariant differentiation of 
a product is:
${(A^K B_L C^M..�)}_{;N} = {A^K}_{;N}(-1)^{[N]([L]+[M]+..)}B_LC^M.. + 
A^K(-1)^{[N]([M]+..)}B_{L;N}C^M..+A^KB_L(-1)^{[N](..)}{C^M}_{;N}..+...$ etc.}

The Riemann curvature tensor arises in the normal way (with suitable
sign factors):
\begin{equation} \label{riemannstart}
(-1)^{[J]}A_J {\cal R}^J{}_{KLM} = A_{K;L;M} - (-1)^{[L][M]} A_{K;M;L}
\end{equation}
Carrying out the algebraic manipulations, we obtain
\begin{eqnarray}
{\cal R}^J{}_{KLM} &=& (-1)^{[J]([K]+[L]+[M])}  \big[(-1)^{[K][L]} 
\Gamma{}_{K M}{}^J{}_{,L} - (-1)^{[K][M]+[L][M]} \Gamma{}_{K L}{}^J{}_{,M}
\nonumber \\
&& \qquad\qquad\qquad\qquad + (-1)^{[L][M]} \Gamma{}_{KM}{}^R \Gamma{}_{R L}{}^J
-\Gamma{}_{KL}{}^R\Gamma{}_{RM}{}^J \big]. \label{riemannformula}
\end{eqnarray}
This is the graded version of the standard Riemann curvature tensor.

It only remains to work out the fully covariant Riemann curvature 
tensor if only to check its graded symmetry properties. Thus we lower
with the metric,
\begin{equation} \label{covriemannstart}
{\cal R}_{JKLM} =  (-1)^{([I]+[J])([K]+[L]+[M])} {\cal R}^{I}{}_{KLM} G_{IJ},
\end{equation}
resulting in
\begin{eqnarray}
{\cal R}_{JKLM}&=&  (-1)^{[J]([K]+[L]+[M])}  \big[(-1)^{[K][L]} 
\Gamma{}_{K M}{}^I{}_{,L} - (-1)^{[K][M]+[L][M]} \Gamma{}_{K L}{}^I{}_{,M} 
\nonumber \\
&&\qquad\qquad\qquad+ 
(-1)^{[L][M]} \Gamma{}_{KM}{}^R \Gamma{}_{RL}{}^I 
-\Gamma{}_{KL}{}^R\Gamma{}_{R M}{}^I \big] G_{IJ}.  \label{covriemannformula}
\end{eqnarray}
It is then not too hard to discover the expected graded symmetry 
relations,
\begin{eqnarray}
{\cal R}_{KJLM} &=& - (-1)^{[J][K]} {\cal R}_{JKLM} , \label{covriemannsym1}\\
{\cal R}_{JKML} &=& - (-1)^{[L][M]} {\cal R}_{JKLM}, \label{covriemannsym2}\\
{\cal R}_{LMJK} &=& (-1)^{([J]+[K])([L]+[M])} {\cal R}_{JKLM}. \label{covriemannsym3}
\end{eqnarray}

\section{\label{sec5}Bianchi identities}
The Riemann curvature tensor also satisfies the Bianchi identities.
the first cyclic identity is readily established from (\ref{covriemannformula})-(\ref{covriemannsym3}):
\begin{equation} \label{bianchi1}
(-1)^{[K][M]} {\cal R}_{JKLM} + (-1)^{[M][L]} {\cal R}_{JMKL} 
+ (-1)^{[L][K]} {\cal R}_{JLMK}  = 0.
\end{equation}
The second (differential) Bianchi identity, involving the covariant 
derivative of the curvature tensor, is most easily uncovered by
proceeding to a ``local frame'' wherein the Christoffel symbol (but
not its derivative) vanishes; in that case the tensor reduces to
${\cal R}_{JKLM;N} = (-1)^{[N]([J]+[K]+[L]+[M])}{\cal R}_{JKLM,N}$. 
With this simplification there emerges the identity
\begin{equation} \label{bianchi2}
(-1)^{[L][N]}{\cal R}_{JKLM;N} +(-1)^{[N][M]}{\cal  R}_{JKNL;M} 
  +(-1)^{ [M][L] }{\cal R}_{JKMN;L} = 0.
\end{equation}

To get the contracted version of the second Bianchi identity, involving
the Ricci tensor, we look at 
\begin{equation} \label{bianchi2constart}
G^{LJ} [(-1)^{[L][N]}{\cal R}_{JKLM;N} +(-1)^{[N][M]}{\cal R}_{JKNL;M}
 +(-1)^{ [M][L] }{\cal R}_{JKMN;L}] = 0.
\end{equation}
This results in
\begin{equation}\label{bianchi2con1}
{\cal R}_{KM;N} - (-1)^{ [N][M]}{\cal R}_{KN;M} +
(-1)^{ [M][L] +[L][N]+[K][L]} G^{LJ} {\cal R}_{JKMN;L} = 0,
\end{equation}
wherein the Ricci tensor has the graded symmetry, 
${\cal R}_{KM}=(-1)^{[M][K]}{\cal R}_{MK}$.
One last contraction with $G^{MK}$ gives
\begin{equation} \label{bianchi2con2}
{\cal R}_{;N} = 2 (-1)^{[M][N]} {\cal R}^{M}{}_{N;M},
\end{equation}
which can be written in the form ${\cal G}^{M}{}_{N;M}=0$ where
\begin{equation}\label{bianchi2con3}
{{\cal G}^M}_N = {{\cal R}^M}_N -{\delta^M}_N{\cal R}/2
\end{equation}
is the graded version of the Einstein tensor. Having established all the
necessary equations with the requisite sign factors, we are in a
position to tackle a simple but important case, featuring one property,
namely charge, and ensuing electromagnetism.

\section{\label{sec6}Gauge changes as property transformations}
To begin tackling the case of one property, an ansatz for the  metric has to 
be made which incorporates the property coordinates. With everything 
flat the metric distance in the manifold $X=(x,\zeta,\bar{\zeta})$ is given by
\[ ds^2 = dX^AdX^B\eta_{BA} = dx^a dx^b\eta_{ba} + 
 d\zeta d\bar{\zeta} \eta_{\bar{\zeta}\zeta}
+d\bar{\zeta}d\zeta  \eta_{\zeta\bar{\zeta}},
 \]
where $\eta_{\zeta\bar{\zeta}}=-\eta_{\bar{\zeta}\zeta}=\ell^2/2$ and
$\eta_{ba}$ is Minkowskian. Notice that we are obliged to introduce 
a fundamental length $\ell$ so as to ensure that the
separation has the correct physical dimensions of length$^2$ because the
$\zeta$ are being taken as dimensionless \footnote{We have chosen to 
use the complex $\bar{\zeta},\zeta$ description rather than the real 
coordinates $\xi,\eta$  (where $\zeta=\xi + i\eta$) because it  lends itself 
more easily to group analysis when one enlarges the number of property
coordinates. }. {\em This should  be construed as the tangent space}. We 
easily spot that it is invariant under Lorentz transformations and global
phase transformations on $\zeta$. However it is not invariant under local 
$x$-dependent phase transformations and we are obliged to introduce the 
gauge field to correct for this, as we shall soon show.

To proceed to curved space we  follow the standard method of invoking
the tetrad formalism, but generalised to a graded space. (The metric is
of course a product of appropriate frame vectors $\cal E$, which provide
the curvature.) We have also been guided by the Kaluza-Klein metric: 
the standard general relativistic metric is to be contained in the 
spacetime-spacetime sector and gauge fields must reside in the 
spacetime-property sector, but we have allowed for U(1) invariant 
property curvature coefficients, denoted by $c_i$.  The results below are
not as adhoc as they may seem; for now we are just interested in seeing
how far one may mimic Klein-Kaluza by using an anticommuting
extension to space-time rather than a commuting one. [The various
terms which arise in the metric below do not allow for fermion contributions
as they would carry a Lorentz spinor index and would conflict with
Lorentz invariance.] We envisage that the $c_i$ are expectation 
values of chargeless Higgs or dilaton fields which ought to be 
considered in the most general situation, left for future research. 

The frame vectors ${{\cal E}_M}^A$ that curve the space
are stated in Appendix 2 and provide the cure for local phase invariance.
They generate the metric in the usual manner \cite{SC,SW} via
\begin{equation} \label{producemetric}
G_{MN} = (-1)^{[B]+[B][N]}{{\cal E}_M}^A \eta_{A B} {{\cal E}_N}^B.
\end{equation}
The entries are tightly constrained by the fact that $G_{mn}$ and
$G_{\zeta\bar{\zeta}}$ have to be bosonic while $G_{m\zeta}$
has to be fermionic in a commutational sense. Further
they only admit expansions up to $\bar{\zeta}\zeta$; that is why the 
electromagnetic field $A_m$ multiplied by $\zeta$ appears in $G_{m\zeta}$;
in principle one could also include in that sector an anticommuting 
ghost field $C_m$ times $\bar{\zeta}\zeta$, as one encounters in 
quantum gravity, but at a semiclassical level we are ignoring this 
aspect of the problem. Putting this all together results in the following metric
\begin{equation} \label{metricform}
G_{MN} = \left(
\begin{array}{ccc}
G_{m n} & G_{m \zeta} & G_{m \bar\zeta}  \\ 
G_{\zeta n} & 0 & G_{\zeta \bar\zeta} \\ 
G_{\bar\zeta n} & G_{\bar\zeta \zeta} & 0
\end{array} \right)
\end{equation}
where
\begin{eqnarray}
G_{m n} &=& g_{mn}(1 + 2 c_1 \bar{\zeta} \zeta)+
e^2\ell^2 A_m A_n \bar{\zeta} \zeta,  \nonumber \\
G_{m \zeta} &=& G_{\zeta m} = -ie\ell^2 A_m \bar{\zeta}/2, \nonumber\\
 G_{m \bar\zeta} &=& G_{\bar\zeta m} =-ie\ell ^2 A_m  \zeta/2, \nonumber \\
G_{\zeta\bar\zeta} &=& -G_{\bar\zeta \zeta}= \ell^2
      (1+2 c_2 \bar{\zeta} \zeta)/2. \label{metricelements}
\end{eqnarray}
A couple of general observations: the charge coupling $e$ 
accompanies the e.m. potential $A$ and the constants $c_i$ are
allowed in the frame vectors to provide phase invariant property 
curvature rather like mass enters the Schwarzschild metric; 
the space-property metric is guaranteed to be anticommuting 
through the factor $\zeta$.

This inverse metric stays graded symmetric, 
$G^{MN} = (-1)^{[M][N]} G^{NM}$, and transforms correctly 
as a rank 2 covariant tensor. It can be derived
from (\ref{metriccon1}) or (\ref{metriccon2}). Its elements are
\begin{eqnarray}
G^{mn} &=& g{}^{m}{}^{n}(1 - 2  c_1\bar{\zeta}\zeta), \nonumber \\
G^{m \zeta} &=& G^{\zeta m} = i eA{}^{m} \zeta,  \nonumber\\
 G^{m \bar\zeta} &=& G^{\bar\zeta m} = - i eA{}^{m} \bar{\zeta}, \nonumber \\
G^{\zeta\bar\zeta} &=& \!\!- G^{\bar\zeta \zeta} = 
   2(1-2 c_2\bar{\zeta}\zeta)/\ell^2\!-\! e^2A^mA_m \bar{\zeta}\zeta. \label{invmetricelements}
\end{eqnarray}

Now suppose that we make a spacetime dependent U(1) 
phase transformation in the property sector:
\begin{equation} \label{propertyphasetrans}
x^\prime = x;  \;\;\;
 \zeta^\prime = e^{i \theta(x)} \zeta; \;\;\;
\bar\zeta^\prime = e^{-i \theta(x)} \bar\zeta.
\end{equation}
Then from the general transformation rules such as (\ref{tensortrans}) and its contravariant
counterpart we readily find that 
\begin{equation} \label{gaugetrans}
eA^\prime{}_m = eA_m + \partial_m \theta,
\end{equation}
which shows the field $A_m$ acts as a gauge field under variations in charge 
phase. This can be checked for all components of the metric $G_{MN}$ from 
the transformation rule (\ref{tensortrans}). On the other hand $G^{mn}$ remains unaffected
and thus is gauge-invariant in the sense of (\ref{propertyphasetrans}) and (\ref{gaugetrans}). 
The same comments apply to ${\cal R}_{mn}$ and ${\cal R}^{mn}$; the former varies 
with gauge but the latter does not.

\section{\label{sec6.5}Metric superdeterminant and Palatini form}
To produce the field equations we require the superdeterminant or Berezinian
of the metric, which is given by  Berezin and DeWitt \cite{BD} to be
\begin{equation} \label{sdetformula}
s\det(X) = \det(A-B D^{-1} C) \det(D)^{-1},
\end{equation}
for a graded matrix of the form:
\begin{equation} \label{sdetmatrix}
X = \left[
\begin{array}{cc}
A & B \\
C & D
\end{array}
\right].
\end{equation}
Given our metric (\ref{metricelements}) this turns out to be:
\begin{equation}\label{sdetG}
s\det(G_{MN}) =  \frac{4}{\ell^4} \det(g{}_{m}{}_{n}) \left[1 + 
(8 c_1-4c_2)\bar{\zeta} \zeta\right]
\end{equation}
or for short,
\begin{equation}\label{sqrtsdetG}
\sqrt{-G..} =  \frac{2}{\ell^2}  \sqrt{-g..} 
\left[1 + ( 4c_1-2c_2) \bar{\zeta} \zeta\right].
\end{equation}
The absence of the gauge potential should be noted. 

While on the subject of the super-determinant we note that  in general, 
$(\sqrt{G..})_{,M}=\sqrt{G..}(-1)^{[N]}{\Gamma_{MN}}^N$, and
not just for our particular $G$. As a direct consequence,
$[\sqrt{G..}A^M]_{;M} = [\sqrt{G..}(-1)^{[M]} {A^M}]_{,M}$, which
impacts on the graded Gauss' theorem. Further, using the
derivative identity,
\begin{equation}
{G^{LK}}_{,M}=-(-1)^{[M][L]}G^{LN}{\Gamma_{NM}}^K
 \!-(-1)^{[K]([N]+[L])}G^{KN}{\Gamma_{NM}}^L,
\end{equation}
we can establish a useful lemma:
\begin{eqnarray}
[\sqrt{G..}G^{MK}]_{,L}\!\!&=&\!(-1)^{[N]}\sqrt{G..}{\Gamma_{LN}}^NG^{MK}-
  \sqrt{G..}[(-1)^{[L][M]}G^{MN}{\Gamma_{NL}}^K\nonumber \\
   & &\qquad +(-1)^{[K]([L]+[M])}     G^{KN}{\Gamma_{NL}}^M]
\end{eqnarray}
so that $[\sqrt{G..}G^{LK}]_{,L}=-\sqrt{G..}G^{LM}{\Gamma_{ML}}^K$ quite simply.
Then because, under an integral sign, the total derivative terms
$[(-1)^{[L]}\sqrt{G..} G^{MK}{\Gamma_{KM}}^L]_{,L},$ and
 $[(-1)^{[L]}\sqrt{G..} G^{MK}{\Gamma_{KL}}^L]_{,M}$ both effectively give zero,
 one can show that
\begin{equation}
 \sqrt{G..}(-1)^{[L]}G^{MK}\left[ (-1)^{[L]([M]+[K])} ({\Gamma_{KM}}^L)_{,L} 
     - ({\Gamma_{KL}}^L)_{,M} \right]  
  \end{equation}
  \begin{equation}\qquad\qquad\qquad   
     =2(-1)^{[L]}\sqrt{G..}G^{MK}\left[
     {\Gamma_{KL}}^N{\Gamma_{NM}}^L-{\Gamma_{KM}}^N{\Gamma_{NL}}^L \right].
\end{equation}
This means that sum of the first two (double) derivative terms in $\sqrt{G..}\cal R$ is
exactly double the sum of the last two terms, apart from a sign change; in other words,
the scalar curvature can be reduced to Palatini form, even in the graded case:
\begin{equation}
\sqrt{G..}{\cal R} \rightarrow (-1)^{[L]}\sqrt{G..}G^{MK}\left[ (-1)^{[L][M]}
  {\Gamma_{KL}}^N{\Gamma_{NM}}^L-{\Gamma_{KM}}^N{\Gamma_{NL}}^L \right].
\end{equation}
This can help to simplify some of the calculations and it also endorses the 
correctness of all our graded sign factors.

\section{\label{sec7}The Ricci Tensor and Superscalar Curvature}
From Eqn (\ref{gammaformula}) and the metric given in (\ref{metricelements}) one may  calculate the Christoffel 
symbols, $\Gamma_{MN}{}^{K}$. A list of these can be found in Appendix 1. 
Using these connections in (\ref{covriemannformula}) one may determine the fully covariant 
Riemann curvature tensor, ${\cal R}_{JKLM}$.  This can be a painful process 
and is where an algebraic computer program developed by one of us (PDS) 
comes in handy, for it minimises the possibility of errors. Even after making use 
of it and the symmetry properties of ${\cal R}$ there are a large number of 
components. We have not bothered to list them as they are so numerous 
and not particularly enlightening. However the contracted Ricci tensor,
\begin{equation} \label{riccistart}
{\cal R}_{KM}= (-1)^{[K][L]} G^{LJ}{\cal R}_{JKLM},
\end{equation}
has fewer entries so we have provided a list of them \footnote{We are reasonably 
certain that those expressions, though complicated,  are correct because we 
have checked that the differential Bianchi identity (\ref{bianchi2con2}) is
 obeyed and this is a highly nontrivial test.} and their contravariant  counterparts,
\begin{equation} \label{conricci}
{\cal R}^{JL}  = (-1)^{[M]} G^{JK} {\cal R}_{KM} G^{ML} ,
\end{equation}
in Appendix 2; the latter are gauge invariant.
Finally the Ricci superscalar can be found by contraction with the metric,
\begin{equation}\label{rdef}
{\cal R} = G^{MK} {\cal R}_{KM}.
\end{equation}

In a frame that is locally flat in spacetime, the spacetime component of the 
contravariant Ricci tensor reduces to
\begin{equation} \label{riccispacetime}
{\cal R}^{mn} \!\!=4g^{mn}c_1[1+(2c_2-6c_1)
\bar{\zeta}\zeta]/\ell^2\!\!-e^2\ell^2 F^{m}{}^{l} F^{n}{}_{l} \bar{\zeta}\zeta/2,
\end{equation}
and the curvature superscalar collapses to
\begin{equation} \label{rscalar}
{\cal R}= 8[4c_1\!-\!3c_2\!+\!c_1(8c_2\!-\!10c_1)\bar{\zeta}\zeta]
/\ell^2 - e^2\ell^2 F^{nl}F_{nl}\bar{\zeta}\zeta/4.
\end{equation}
Both expressions (\ref{riccispacetime}) and (\ref{rscalar})  are gauge independent. By making 
use of them and the superdeterminant (\ref{sqrtsdetG}) we may evaluate firstly
the total Lagrangian density for electromagnetic property,
\[
{\cal L}=\int\!\!d \zeta d \bar\zeta \sqrt{-G..}\,{\cal R}
     \propto -\frac{1}{4}F_{mn}F^{mn} + \frac{48(c_1-c_2)^2}{e^2\ell^4},
\]
and secondly the Einstein tensor in flat spacetime:
\[
\int\!\!\!d \zeta d \bar\zeta \sqrt{-G..}({\cal R}^{km}-{\cal R}\; G^{km}/2)\!\!\propto\!\!
\left[ 48c_2(c_1\!\!-\!\!c_2)g^{km}/e^2\ell^4
\!-\!( {F^k}{}^{l} F^{m}{}_{l} -F_{ln} F^{ln} g^{km}/4)\right].
\]
The familiar expression for the electromagnetic stress tensor, 
namely $T^{km}\equiv  {F^k}_l F^{lm} + F_{nl}F^{nl}g^{km}/4$
emerges naturally and becomes part of the geometry. But we also 
recognize a cosmological constant term that  is largely determined by 
the magnitude $(c_2-c_1)/\ell^4$.
[As an aside, we have verified that (\ref{riccispacetime}) and (\ref{rscalar}), 
remain true in a general frame, not necessarily locally flat.]

Including gravity by curving spacetime means including the standard 
gravitational curvature $R$ and will render (\ref{riccispacetime}) and (\ref{rscalar}) generally 
covariant. (One has be careful here to track factors of $\bar{\zeta}\zeta$,
as one will be integrating over property.)  It straightforward to see
that the gravitational part of the superscalar ${\cal R}^{(g)}$ is 
$R(1-2c_1\bar{\zeta}\zeta)$, while the super-Ricci tensor
${\cal R}^{(g)km}$ contains $R^{km}(1-4c_1\bar{\zeta}\zeta)$. In
consequence we may evaluate the full gravitational-electromagnetic 
Lagrangian through the property integral:
\begin{equation} \label{lagrangian}
{\cal L}=\int\! d \zeta d \bar\zeta \sqrt{-G..}\,{\cal R}
   =2e^2\sqrt{-g..} \left[
     \frac{2(c_1-c_2)R}{e^2\ell^2} -\frac{F_{mn}F^{mn}}{4} +
  \frac{48(c_1-c_2)^2}{e^2\ell^4}\right],
\end{equation}
wherein we recognize 
$$16\pi G_N\equiv \kappa^2 = e^2\ell^2/2(c_1-c_2),
\quad \Lambda = 12(c_2-c_1)/\ell^2.$$
To verify that the entire setup is consistent and free of error we may
determine the gravitational variation $\delta G_{MN}$ which equals
$\delta g_{mn}(1+2c_1\bar{\zeta}\zeta)$. Hence the gravitational field
equation is obtained through
\begin{eqnarray} \label{einsteintensor}
0=\int d\zeta d\bar{\zeta} \sqrt{-G..}&(&\!\!\!\!\!1+2c_1\bar{\zeta}\zeta)({\cal R}^{km} 
-G^{km}{\cal R} / 2 )
\nonumber \\
&=&\!\!\!\sqrt{-g..}\left[\frac{4(c_1\!\!-\!\!c_2)}{\ell^2}(R^{km}\!\!-\!\frac{1}{2}g^{km}R)
 \!-\!  T^{km}\! - \!\frac{48(c_1\!\!-\!c_2)^2}{\ell^4}g^{km}\!\right].
\end{eqnarray}
This is just what we would have obtained from (50). In any case we see
that the universal coupling of gravity to stress tensors $T$ has a factor
$8\pi G_N\equiv \kappa^2/2 = e^2\ell^2/4(c_1-c_2) > 0$.  The result is
to make the cosmological term go negative and, what is probably worse,
it has a value which is inordinately larger than the tiny experimental 
value found by analyses of supernovae! (All cosmological terms derived 
from particle physics, except for exactly zero, share the same problem). 
Numerically speaking, $\kappa\simeq 5.8\times 10^{-19}$ (GeV)$^{-1}$ 
means $\ell \sim 10^{-18}$ (GeV)$^{-1}$ is Planckian in scale. Of course
the magnitude of the miniscule cosmological constant $\Lambda\sim4\times
10^{-84}$ (GeV)$^2$  is at variance with Planckian expectations by the usual 
factor of $10^{-120}$, which is probably the most mysterious natural ratio. So 
far as our scheme is concerned, we are disappointed but not particularly 
troubled by the wrong sign of $\Lambda$ because it can readily be 
reversed by  extra property curvature coefficients when we enlarge the 
number of properties (as we have checked when enlarging the 
number of properties to at least two). The magnitude of $\Lambda$ is 
quite another matter because it will require some extraordinary fine-tuning, 
even after fixing the sign.

\section{\label{sec8}Inclusion of matter fields}
The conventional results which we obtained for electromagnetism plus 
gravity, through the property of electricity, merely confirm the fact that 
our scheme is perfectly viable and offers a novel perspective on nature. 
We anticipate that when one incorporates other 
properties, like chromicity and neutrinicity, then the usual picture of QCD
plus gravity plus electroweak theory will emerge. For now we wish to
exhibit some preliminary research concerning inclusion of matter fields, 
despite being limited to the single property of electric charge. 

\subsection{Scalar Field}

Adhering to the tenets of the spin-statistics connection, we begin by assuming 
that a superscalar field $\Phi(X)$ is overall Bose and can be expanded into
even powers of $\bar{\zeta}\zeta$; thus it has the general form 
$\Phi(x,\zeta,\bar{\zeta}) = U(x) + V(x)\bar{\zeta}\zeta$. Note that we could have 
included in $\Phi$ two anticommuting scalar ghost fields in the combination
$\bar{\zeta}C + \bar{C}\zeta$; such ghost fields have a place in quantum 
theory but, with their incorrect spin-statistics, cannot be regarded as
physical asymptotic states. The same comment applies to the 
spacetime-property sector where we could have included vector ghost fields
of the type $C_m, \bar{C}_m$ multiplying $(1+c'\bar{\zeta}\zeta)$. We have
ignored these extras as we are only dealing with semiclassical e.m./gravity
for the purpose of the present investigation, but they are sure to come in 
their own when quantization of the scheme is undertaken. By imposing 
self-duality so $U(x)=V(x)=\varphi(x)$, $\Phi$ may be reduced to the 
form \cite{RDPJRW,RD4}
\begin{equation} \label{superscalarfield}
\Phi(X) = \varphi(x) (1+\bar{\zeta}\zeta)/2.
\end{equation}
Necessarily $\varphi$ carries zero charge and is a far cry from a Higgs
field. (In fact to obtain the correct quantum numbers of the Higgs field it
is imperative to attach three chromic properties to charge.) As we will be 
coupling this field to the supermetric, which brings in the superdeterminant
$2\sqrt{-G..}/\ell^2$, we shall introduce an extra factor of $\ell^2$
to eliminate this scale and we will also ignore $c_i$ curvature in what 
follows except from what the mixed $x-\zeta$ sector produces. 

A mass term in the Lagrangian of $\mu^2\varphi^2/2$ will arise through the 
property integral
\begin{equation} \label{masstermorigin}
(\ell^2/2)\!\!\int\!\!d\zeta d\bar{\zeta}\sqrt{-G..}\,\mu^2\Phi^2 = \!\!
\int\!\!d\zeta d\bar{\zeta}\sqrt{-g..}(1+2\bar{\zeta}\zeta)\mu^2\varphi^2/4.  
\end{equation}
The kinetic term is more interesting, because it adds to the mass
owing to the property components in $\Phi$; specifically this can be
attributed to the $\zeta,\bar{\zeta}$ derivatives, which add a piece to the 
mass.  Thus we consider
\begin{equation} \label{kineticterm}
(\ell^2/2)\!\!\int\!\!d\zeta d\bar{\zeta}\,\, \sqrt{-G..}\,\, G^{MN}
\partial_N\Phi\,\partial_M\Phi.
\end{equation}
Upon inserting the metric from (\ref{metricelements}), we find that the contributions from the 
gauge field cancel out, {\em as they must}, and we are left with
\begin{equation}\int d\zeta d\bar{\zeta}\sqrt{-g..}[(1+2\bar{\zeta}\zeta)\,g^{mn}\partial_n\phi
    \,\partial_m\phi/4+\bar{\zeta}\zeta\varphi^2/\ell^2].
\end{equation}
 The only feasible way then to cancel off mass in order to obtain a massless
 scalar field in this scheme is to match the $\varphi^2/\ell^2$ from the 
 property  kinetic energy to the previously constructed mass term (\ref{masstermorigin}).

\subsection{Spinor Field}
In seeking a generalisation of the Dirac equation to incorporate the graded
derivative, we need to bear in mind that the electromagnetic potential
is embedded in the spacetime-property frame vector ${E_A}^M$; therefore
we first need to determine the inverse vielbein, obtained via the condition
${{\cal E}_M}^A {E_A}^N = {\delta_M}^N$. The components are listed
in the second appendix, where it will be seen that the vector potential
is held in the space-property sector  via ${E_a}^\zeta$ and 
${E_a}^{\bar{\zeta}}$. Since the Dirac operator has a natural extension from
$i\gamma^a{e_a}^m\partial_m$ to $i\Gamma^A{E_A}^M\partial_M$, it is 
vital to include the graded derivative $\partial/\partial\zeta$ at the 
very least.

Just as Dirac was obliged to enlarge spinors from two to four components
in order to go from non-relativistic electrons to relativistic ones, so too we are 
forced to extend the space in order to deal with the property derivatives.
(There may be other ways to attain that goal.)
Since the Dirac operator will act on a spinorial superfield, we have been 
led to consider an extended field $\Psi(X) = \theta\bar{\zeta}\psi(x)$
and the representation:
\begin{equation}\label{superdirac}
 \Gamma^a = \gamma^a,\, \ell\Gamma^\zeta=2i\partial/\partial\theta,
 \, \ell\Gamma^{\bar{\zeta}} = 2i\theta,
\end{equation}
wherein $\theta$ is {\em another} complex scalar a-number and we
eventually have to integrate over $\theta$ and $\bar{\theta}$. (The
$\Gamma^{\zeta},\Gamma^{\bar{\zeta}}$ act like fermionic annihilators).
The action of the extended Dirac operator then yields
\begin{eqnarray}
i\Gamma^A{E_A}^M\partial_M\Psi &=&\left[i\gamma^a{e_a}^m\partial_m+
e\gamma^aA_a\bar{\zeta}\frac{\partial}{\partial{\bar{\zeta}}}
+\,\,\frac{2}{\ell}(1-f\bar{\zeta}\zeta)\frac{\partial^2}
{d\theta d\bar{\zeta}} \right] \theta\bar{\zeta}\psi\nonumber\\
&=&\theta\bar{\zeta}
\gamma^a{e_a}^m(i\partial_m\! 
 \!+\!eA_m)\psi \!-\!\!\frac{2}{\ell}(1\!-\!\!f\bar{\zeta}{\zeta})\psi.\label{extdirac}
\end{eqnarray}
This means that when we include the adjoint $\bar{\Psi}\equiv 
-\bar{\psi}\zeta\bar{\theta}$ and integrate over the subsidiary $\theta,\zeta$,
we end up with the normal gauge invariant spinorial Lagrangian density:
\begin{eqnarray}
{\cal L}&=&\!\!\!\int(d\zeta d\bar{\zeta})(d\theta d\bar{\theta})\bar{\Psi}(X)
 \left[i\Gamma^A{E_A}^M\partial_M - {\cal M}\right] \!\!\Psi(X)\nonumber\\
 &=& \bar{\psi}(x)\left[\gamma^a {e_a}^m(i\partial_m+eA_m) \label{spinlagrangian}
 -{\cal M}\right]\psi(x).
\end{eqnarray}
Very likely there exists a more elegant way of reaching (\ref{spinlagrangian}) but
however this is done the coupling of the charged fermion to the
spacetime-property vielbein, which contains $A$, is critical. The 
representation (\ref{superdirac}) will surely need revisiting in order to encompass 
chirality if the basic fermions are taken as left-handed, especially if we 
attach charge conjugate left-handed pieces, in order to encompass all 
spin states.

\section{\label{sec9}Conclusions}
The framework underlying our research was inspired by supersymmetry, but
instead of using auxiliary {\em spinor} coordinates we have made them 
{\em scalar} and connected them with something {\em tangible}, namely 
property or attribute. This point is important because all physical events are 
described by changes in momentum and/or property. 
From this perspective, systematisation of property, with the natural 
occurrence of generations, becomes a guiding principle. 
An obvious criticism of the approach is that the superstructure has only 
led to the standard Einstein-Maxwell Lagrangian, which is hardly an 
earth-shattering conclusion! True, but by geometrizing spacetime-property
we have succeeded in reinterpreting gauge fields as the messengers of 
property in a larger graded curved space, besides offering a new 
viewpoint on the nature of events; as a bonus we see that curvature in
property space can act as a source of a cosmological constant --- this with
 just one property coordinate -- even if its value is ridiculous. Furthermore
 addition of further $\zeta$ coordinates offers a natural path to group theory
 classification \footnote{To stress this point we mention that we have
 succeeded in obtaining the combined Yang-Mills-Gravity Lagrangian
 for two property coordinates. The details are much more intricate than the
 one $\zeta$ case considered in this paper and will be submitted for
 publication separately.} without entraining infinite towers of states 
 as one gets with  bosonic  extensions to spacetime.

We foresee no intrinsic difficulty in extending the work to QCD or to 
electroweak theory, though the algebraic manipulations will perforce
be more intricate. Going all the way, we anticipate an extension of our 
calculations to five property coordinates ($\zeta^0$ to $\zeta^4$) and the 
distillation of the final group to that of the standard model will mean that 
property curvature coefficients $c_{\rm col}, c_{\rm ew}$ are to be associated 
with colour and electroweak invariants, $\bar{\zeta}_i\zeta^i$ and 
$(\bar{\zeta}_0\zeta^0+\bar{\zeta}_4\zeta^4)$ respectively --- perhaps 
engendered by expectation values of chargeless Higgs or dilaton fields. 
For the future these are just the most prominent issues that come to mind
with correction of the $\Lambda$ sign foremost among them and the 
quantization of the scheme in the present framework as the next step.
There are surely several research avenues \footnote{Looking beyond the horizon
of this paper, since the gravitational coupling  to any stress tensor is universal  
and since we have married couplings with fields 
 it means that these interaction couplings will need to be universal too. 
 At low energies QCD and QED coupling constants $e$ and $g$ are widely different 
 so, unless the curvature coefficients $c_e$ and $c_w$ are taken to be different which is
 entirely possible,  we  envisage a scenario where these couplings are running and 
 unified  at a GUT-like scale $\ell$ with $1/\alpha\simeq 40$; they only look different 
 when we run down from $e^2(\ell^2) = g^2(\ell^2)$ to electroweak/strong 
 interaction scales.} to explore in the present picture, many more than we have 
 envisaged. 

In the end, this geometrical approach of uniting gravity with other natural 
forces through a larger graded space-time-property manifold may turn out to 
be quite misguided. That would be  disappointing as it is hard to imagine what
other original way one could unify the gravitational field with other fields. 
As a fallback position we 
could be ultraprudent  by abandoning the unification goal: just introduce 
gauge fields in the time-honoured way, ensuring that differentiation is 
gauge covariant under property transformations, by replacing ordinary 
derivatives $\partial$ with $D =\partial + ieA.T$, where the generators 
$T$ are represented by  property rotations acting on matter superfields; 
e.g. the charge operator by $(\bar{\zeta}\,\partial/\partial\bar{\zeta} - 
\zeta\,\partial/\partial\zeta).$ That would be a backward step and even then, 
we might be confronted by insurmountable obstacles. However, for now, the 
geometric scheme seems flexible enough and accords well with our general 
understanding of fundamental particle physics and its content as well as 
gravitation. Should our framework fall by the wayside, there is no other 
recourse than to persist with variants of  grand theories which are currently 
on the market, in the hope that experiment will, at sufficiently high energy, 
substantiate one of  them. Failing that, we trust that one day somebody 
will conceive a radically new description of events that will lead to new 
insights with {\em testable} predictions.

\section{Acknowledgments}
We would like to thank Dr Peter Jarvis for much helpful advice 
and for his knowledgeable comments about supergroups, their 
representations and their dimensions.

\appendix
\section{Christoffel symbols, Vielbeins and Curvature components}
\subsection{The graded Christoffel connections}
From definition (\ref{gammaformula}) and the metric elements 
(\ref{metricelements}) one may derive the following components
of the Christoffel symbols:
\begin{eqnarray*}
\Gamma_{m n}{}^{l} &=&  \Gamma^{[g]}{}_{m n}{}^{l} + e^2\ell^2 (A{}_{n} F{}_{m}{}_{k}
+A{}_{m} F{}_{n}{}_{k}) g{}^{k}{}^{l} \bar{\zeta}\zeta/2,
\\
\Gamma_{m n}{}^{\zeta} \!\!&=& \!\!\frac{\zeta}{2} \bigg[ie\big(2A{}^{k} \Gamma^{[g]}_{m n k}
\!-\! A{}_{m}{}_{,n}\!-\! A{}_{n}{}_{,m} \big) \!\!-\!\!2e^2A{}_{m} A{}_{n}\!\!- \!\!\frac{4c_1}{\ell^2}  
g{}_{m}{}_{n}\! \bigg],
\\
\Gamma_{m n}{}^{\bar\zeta}\!\!& =&\!\! \frac{\bar{\zeta}}{2} \bigg[ ie 
\big(A{}_{m}{}_{,n}\!+\! A{}_{n}{}_{,m} 
\!-\!2A{}^{k} \Gamma^{[g]}_{m n k}\big)\!\!-\!\!2e^2\!A{}_{m} A{}_{n}
\!\!-\!\!\frac{4 c_1}{\ell^2} g{}_{m}{}_{n} \!\bigg],
\\
\Gamma_{\zeta n}{}^{l}&=&\! \Gamma_{n \zeta}{}^{l}\!=\!\bar{\zeta} \bigg[
ie\ell ^2 F_{kn} g{}^{k}{}^{l}/4-c_1\delta_{n}{}^{l}\bigg],
\\
\Gamma_{\bar\zeta n}{}^{l}&=& \Gamma_{n \bar\zeta}{}^{l}\!=\!\zeta \bigg[
ie\ell^2 F_{k n} g{}^{k}{}^{l}/4+c_1 \delta_{n}{}^{l}\bigg],
\\
\Gamma_{\zeta n}{}^{\zeta}\!&=&\!\Gamma_{n \zeta}{}^{\zeta}\!=\!
-ieA{}_{n}-\!\bigg[\frac{e^2\ell^2}{4} A{}^{k} F_{k n}+ ie(c_1-2c_2)A{}_{n}\!
 \bigg] \!\bar{\zeta}\zeta,
\\
\Gamma_{\zeta n}{}^{\bar\zeta}&=& \Gamma_{n \zeta}{}^{\bar\zeta} = 
\Gamma_{\bar\zeta n}{}^{\zeta}= \Gamma_{ n \bar\zeta}{}^{\zeta}=  0,
\\
\Gamma_{\bar\zeta n}{}^{\bar\zeta}&=& \Gamma_{n \bar\zeta}{}^{\bar\zeta}=
ieA{}_{n}-\bigg[ \frac{e^2\ell^2}{4}A{}^{k} F_{kn}+ie(2c_2 -c_1)A{}_{n}\!\bigg] 
\!\bar{\zeta}\zeta,
\\
\Gamma_{\zeta \bar\zeta}{}^{l}&=& \Gamma_{ \bar\zeta \zeta}{}^{l} = 0,
\\
\Gamma_{\zeta \bar\zeta}{}^{\zeta}&=& -\Gamma_{\bar\zeta \zeta}{}^{\zeta}  
= -2 c_2 \zeta,\\
\Gamma_{\zeta \bar\zeta}{}^{\bar\zeta}&=& -\Gamma_{\bar\zeta \zeta }{}^{\bar\zeta}= 
-2 c_2\bar{\zeta},\\
\Gamma_{\zeta \zeta}{}^l &=&\Gamma_{\zeta \zeta}{}^\zeta = 
\Gamma_{\zeta \zeta}{}^{\bar\zeta} = \Gamma_{\bar\zeta \bar\zeta}{}^l = 
\Gamma_{\bar\zeta \bar\zeta}{}^\zeta = \Gamma_{\bar\zeta \bar\zeta}{}^{\bar\zeta} = 0.
\end{eqnarray*}
Above, $\Gamma^{(g)}$ signifies the purely gravitational connection and
$F_{mn}\equiv A_{n,m}-A_{m,n}$ is the standard Maxwell tensor.
These connections are essential in determining the full super-Riemann components.

\subsection{Ricci Tensor Components}
A concise list of the Ricci supertensor components (contravariant and covariant) is
as follows, where we neglect spacetime curvature:
\begin{eqnarray*}
{\cal R}^{km} \!\!&=&4g^{mk}c_1[1+(2c_2-6c_1)\bar{\zeta}\zeta]/\ell^2
-e^2\ell^2  F^{k}{}_{l} F^{m}{}^{l}\bar{\zeta}\zeta/2,\\
{\cal R}^{k \zeta}&=&  4 ie c_1 A^k \zeta/\ell^2-ieF^{kl}{}_{,l}\zeta/2,\\
{\cal R}^{k \bar\zeta} &=&-4iec_1\bar{\zeta}A{}^{k} /\ell^2+ie \bar{\zeta} F^{kl} {}_{,l}/2,\\
{\cal R}^{\zeta \bar\zeta}&=&8[3c_2(1\!-\!2c_2 \bar{\zeta}\zeta)
-2c_1(1\!-\!c_1\bar{\zeta}\zeta)]/\ell^4\\
& &-e^2(4 A^mA_m c_1/\ell^2 +F_{mn} F^{mn}/4
+\!A_m {F^{nm}}_{,n} )\bar{\zeta}\zeta,\\
{\cal R}_{km} &=&4c_1 g_{km}[1-2(c_1\!-\!c_2)\bar{\zeta}\zeta]/\ell^2
-4e^2(2c_1-3 c_2) A_k A_m \bar{\zeta}\zeta\\
& &+e^2\ell^2g^{nl}\big[ A_{k,n}A_{m,l}-A_{n,m}A_{l,k}\big ] \bar{\zeta}\zeta/2,\\
{\cal R}_{k \zeta} &=& 2ie (2c_1-3c_2) \bar{\zeta} A{}_{k}+ie\ell^2 \bar{\zeta} {F^l}_{k,l}/4,\\
{\cal R}_{k \bar\zeta}&=&2ie(2c_1-3c_2)A_k\zeta +ie\zeta \ell^2 {F^l}_{k,l}/4,\\
{\cal R}_{\zeta \bar\zeta} &=& [6c_2-4c_1 + 4(c_2-c_1)(3c_2-c_1)
\bar{\zeta}\zeta ]- \ell^4 e^2F^{kl}F_{kl}\,\bar{\zeta}\zeta/16.
\end{eqnarray*}
Other components are derivable from symmetry properties of ${\cal R}_{MN}$.

\subsection{Vielbeins}
From the frame vectors, namely ${{\cal E}_M}^A$, whose components  are
\begin{eqnarray*}
{{\cal E}_m}^a\!\!\!&=&\!\!\!(1\!\!+\!c_1\bar{\zeta}\zeta){e_m}^a,\,\,\,
 {{\cal E}_m}^\zeta\!\!=\!\!-ie\bar{\zeta}A_m,\,\,\,
 {{\cal E}_m}^{\bar{\zeta}}\!\!=\! -ieA_m\zeta,\nonumber\\
 {{\cal E}_\zeta}^a &=& 0, \quad{{\cal E}_\zeta}^\zeta=0,\quad
 {{\cal E}_\zeta}^{\bar{\zeta}}=(1+c_2\bar{\zeta}\zeta),\nonumber\\
 {{\cal E}_{\bar{\zeta}}}^a &=&0,
 \quad {{\cal E}_{\bar{\zeta}}}^\zeta = -(1+c_2\bar{\zeta}\zeta),\quad
  {{\cal E}_{\bar{\zeta}}}^{\bar{\zeta}} = 0,
  \end{eqnarray*}  
\noindent we may derive the super-vielbeins ${E_A}^N$, obtained via 
${{\cal E}_M}^A {E_A}^N = {\delta_M}^N$. In this way we arrive at the set:
\begin{eqnarray*}
 {E_a}^m\!&=&\!\!{e_a}^m(1\!-\!c_1\bar{\zeta}\zeta),\,\,\,
 {E_a}^\zeta\!=\!ieA_a\zeta,\,\,\,
 {E_a}^{\bar{\zeta}}\!=\!-ie\bar{\zeta}A_a,\nonumber\\
 {E_\zeta}^m &=& 0,\quad {E_\zeta}^\zeta = 0,\quad
 {E_\zeta}^{\bar{\zeta}} = -(1-c_2\bar{\zeta}\zeta),\nonumber\\
 {E_{\bar{\zeta}}}^m&=&0,\quad  {E_{\bar{\zeta}}}^\zeta=(1-c_2\bar{\zeta}\zeta),
 \quad   {E_{\bar{\zeta}}}^{\bar{\zeta}} = 0.
\end{eqnarray*}
These expressions are required in Section \ref{sec8} B. As a useful check on their
correctness we may ascertain that $G^{MN}= (-1)^{[A][M]}\eta^{AB}
{E_B}^M {E_A}^N$, emerges properly. For example we directly arrive at
\begin{eqnarray*}
G^{\zeta\bar{\zeta}} &=&\eta^{ab} {E_b}^\zeta {E_a}^{\bar{\zeta}} -
2{E_{\bar{\zeta}}}^\zeta {E_\zeta}^{\bar{\zeta}}/\ell^2
+2{E_\zeta}^\zeta {E_{\bar{\zeta}}}^{\bar{\zeta}}/\ell^2\\
&=&\eta^{mn}(ieA_n\zeta)(-i\bar{\zeta}eA_m)
+2(1-2c_2\bar{\zeta}\zeta)/\ell^2\\
&=&2(1-2c_2\bar{\zeta}\zeta)/\ell^2-e^2A^mA_m\bar{\zeta}\zeta
\end{eqnarray*}


\vspace{1cm}
\begin{thebibliography}{99}
\bibitem{BD} F.A.~Berezin,  \emph{The method of second quantization}, (Academic
Press, Boston, 1966);
\bibitem{BDW} B.~DeWitt, \emph{Supermanifolds}, (Cambridge University Press,
Cambridge, 1984).
\bibitem{FB} F.A.~Berezin, \emph{Introduction to Superanalysis}, 
(D.Reidel Pub. Co., Dordrecht, 1987).
\bibitem{AR} A.~Rogers, \emph{Supermanifolds: Theory and Applications},
(World Scientific, Singapore, 2007).
\bibitem{EW} E.~Witten, \emph{Notes on Supermanifolds and Integration},
{\bf arXiv}:1209.2199v2 (2102).
\bibitem{RDPJRW} R.~Delbourgo, P.D.~Jarvis and R.~Warner, J. Math. Phys. 
{\bf 34}, 3616 (1993)
\bibitem{RDRZ} R.~Delbourgo and R.B.~Zhang, Phys. Rev. {\bf D38}:2490 (1988).
\bibitem{RD2} R.~Delbourgo, J. Phys. {\bf A39}, 5175 (2006); {\em ibid}, 14735 (2006)
\bibitem{RD3} R.~Delbourgo, Int. J. Mod. Phys. {\bf 22A}, 29 (2007).
\bibitem{RD4} R.~Delbourgo, arXiv:1202.4216 (2012).
\bibitem{RDPDJGT} R.~Delbourgo, P.D.~Jarvis and G.~Thompson, Phys. Lett.
 {\bf B109}, 25 (1982)
 \bibitem{RDPDJGT2} R.~Delbourgo, P.D.~Jarvis and G.~Thompson,
 Phys. Rev. {\bf D26}, 775 (1982).
\bibitem{MAPL} M.~Asorey and P.M.~Lavrov, J. Math. Phys, {\bf 50}:013530 (2009).
\bibitem{SC} S.~Carroll, \emph{Spacetime and Geometry}, (Addison Wesley, San Francisco, 2004)
\bibitem{SW} S.~Weinberg, \emph{Gravitation and Cosmology}, (J.Wiley \& Sons, New York, 1972)
 \end{thebibliography}
\end{document}